\documentclass[12pt]{article}
\usepackage{psfrag}
\usepackage{epsfig}
\usepackage{amsmath}
\usepackage{hhline}
\usepackage{amssymb}
\usepackage{times}
\usepackage{cite}
\newlength{\dinwidth}
\newlength{\dinmargin}
\newcommand{\jpsi}{$J/\psi$}

\newcommand{\psits}{$\psi(2S)$}

\newcommand{\wgp}{$W_{\gamma p}$}
\newcommand{\wgpw}{W_{\gamma p}}

\newcommand{\degree}{\rm \circ}
\newcommand{\GeV}{\,\mbox{GeV}}

\newcommand{\gev}{\,\mbox{GeV}}

\begin{document}  
\newcommand{\qsq}{\ensuremath{Q^2} }
\newcommand{\gevsq}{\ensuremath{\mathrm{GeV}^2} }
\newcommand{\et}{\ensuremath{E_t^*} }
\newcommand{\rap}{\ensuremath{\eta^*} }
\newcommand{\gp}{\ensuremath{\gamma^*}p }
\newcommand{\dsiget}{\ensuremath{{\rm d}\sigma_{ep}/{\rm d}E_t^*} }
\newcommand{\dsigrap}{\ensuremath{{\rm d}\sigma_{ep}/{\rm d}\eta^*} }
\newcommand{\ra}{\rightarrow}
\newcommand{\psip}{$\psi(2S)$}
\newcommand{\ccbar}{\ensuremath{\mit c\overline{c}}}

\def\Journal#1#2#3#4{{#1} {\bf #2} (#3) #4}
\def\NCA{\em Nuovo Cimento}
\def\NIM{\em Nucl. Instr. Methods}
\def\NIMA{{\em Nucl. Instr. Methods} {\bf A}}
\def\NPB{{\em Nucl. Phys.}   {\bf B}}
\def\PLB{{\em Phys. Lett.}   {\bf B}}
\def\PRL{\em Phys. Rev. Lett.}
\def\PRD{{\em Phys. Rev.}    {\bf D}}
\def\ZPC{{\em Z. Phys.}      {\bf C}}
\def\EJC{{\em Eur. Phys. J.} {\bf C}}
\def\CPC{\em Comp. Phys. Commun.}

\begin{titlepage}

\noindent
DESY 02-075  \hfill  ISSN 0418-9833 \\
May 2002

\vspace*{3cm}

\begin{center}
\begin{Large}

{\bf Diffractive Photoproduction of {\boldmath \psip} Mesons at HERA}

\vspace*{1cm}

H1 Collaboration

\end{Large}
\end{center}

\vspace{3cm}

\begin{abstract}
Results on diffractive photoproduction
of $\psi(2S)$ mesons are presented using 
data collected between 1996 and 2000
with the H1 detector at the HERA $ep$ collider.
The data \linebreak[4] correspond to an integrated luminosity of 77 pb$^{-1}$.
The energy dependence of the \linebreak[4] diffractive $\psi(2S)$ 
cross section is found to be similar to or possibly somewhat steeper than
that for $J/\psi$ mesons. 
The dependences of the elastic and proton dissociative \linebreak[4]
$\psi(2S)$ photoproduction cross sections on
the squared momentum transfer $t$ at the 
proton vertex are measured.
The $t$-dependence of the elastic channel, parametrised as $e^{bt}$, yields 
$b_{el}^{\psi(2S)}=(4.31\pm0.57\pm0.46)$ GeV$^{-2}$, compatible with that
of the $J/\psi$. For the proton dissociative channel the result
$b_{pd}^{\psi(2S)}=(0.59\pm0.13\pm0.12)$ GeV$^{-2}$ is 2.3 standard deviations
smaller than that measured for the $J/\psi$.
With proper account of the individual wavefunctions theoretical predictions based on perturbative QCD are found to describe the measurements well.
\end{abstract}

\vspace{1.5cm}

\begin{center}
To be submitted to Phys. Lett. B
\end{center}

\end{titlepage}

%
%

\begin{flushleft}


C.~Adloff$^{33}$,              
V.~Andreev$^{24}$,             
B.~Andrieu$^{27}$,             
T.~Anthonis$^{4}$,             
A.~Astvatsatourov$^{35}$,      
A.~Babaev$^{23}$,              
J.~B\"ahr$^{35}$,              
P.~Baranov$^{24}$,             
E.~Barrelet$^{28}$,            
W.~Bartel$^{10}$
% ...

\bigskip{\it
 $ ^{1}$ I. Physikalisches Institut der RWTH, Aachen, Germany$^{ a}$ \\
 $ ^{2}$ III. Physikalisches Institut der RWTH, Aachen, Germany$^{ a}$ \\
% ...
$ ^{4}$ Inter-University Institute for High Energies ULB-VUB, Brussels;
          Universiteit Antwerpen (UIA), Antwerpen; Belgium$^{ c}$ \\
% ...
$ ^{10}$ DESY, Hamburg, Germany \\
% ...
$ ^{23}$ Institute for Theoretical and Experimental Physics,
          Moscow, Russia$^{ l}$ \\
 $ ^{24}$ Lebedev Physical Institute, Moscow, Russia$^{ e}$ \\
% ...
$ ^{27}$ LPNHE, Ecole Polytechnique, IN2P3-CNRS, Palaiseau, France \\
 $ ^{28}$ LPNHE, Universit\'{e}s Paris VI and VII, IN2P3-CNRS,
          Paris, France \\
% ...
$ ^{33}$ Fachbereich Physik, Bergische Universit\"at Gesamthochschule
          Wuppertal, Wuppertal, Germany \\
% ...
$ ^{35}$ DESY, Zeuthen, Germany \\
% ...
$ ^a$ Supported by the Bundesministerium f\"ur Bildung und Forschung, FRG,
      under contract numbers 05 H1 1GUA /1, 05 H1 1PAA /1, 05 H1 1PAB /9,
      05 H1 1PEA /6, 05 H1 1VHA /7 and 05 H1 1VHB /5 \\
% ...
$ ^c$ Supported by FNRS-FWO-Vlaanderen, IISN-IIKW and IWT\\
% ...
$ ^e$ Supported by the Deutsche Forschungsgemeinschaft \\
% ...
$ ^l$ Partially Supported by Russian Foundation
      for Basic Research, grant    no. 00-15-96584 \\
}

\end{flushleft}

\newpage

\section{Introduction}

\noindent
We report on a measurement of diffractive photoproduction 
of $\psi(2S)$ mesons, $\gamma p \rightarrow \psi(2S)+ Y$,
where $Y$ denotes either a proton or a proton dissociation system
of mass $M_Y > m_p$. The data were taken with the H1 Detector 
at the $ep$ collider HERA.
The dependences of the cross section on  
\wgp, the photon-proton center of mass energy, and on $t$, 
the square of the four-momentum transfer at the 
proton vertex, are measured for the first time
and are compared to those of $J/\psi$ production. 
The data cover the region $ 40 < W_{\gamma p} < 150 $ GeV.

\noindent 
In recent years diffractive production of $J/\psi$ mesons has been 
measured at HERA with increasing precision 
\cite{Aid:1996dn,Breitweg:1997rg,Adloff:2000vm,Chekanov:2002xi}. 
The cross section for the elastic process 
$\gamma p \rightarrow J/\psi\,p$ was found to 
rise steeply with \wgp. This was interpreted 
as a signature for a ``hard'' process and calculations in perturbative 
Quantum Chromodynamics (pQCD) were able to give a good description 
of the data \cite{Martin:2000wb,Frankfurt:2001ez}.
In perturbative QCD, diffractive charmonium production in $\gamma p$ scattering
can be viewed in the proton rest frame 
as a sequence of several steps separated in time.
An almost real photon is emitted from the incoming lepton and fluctuates into 
a $c\bar{c}$ pair. The $c\bar{c}$ pair subsequently interacts with 
the proton via the exchange of two gluons
(or a gluon ladder) in a colour-singlet state and 
then evolves into a real vector meson. 
In such models, with the $c\bar{c}$ fluctuation of the photon treated 
as a colour dipole, a number of distinct predictions are made
for \psip\ photoproduction\cite{Kopeliovich:1991pu,Kopeliovich:1993gk,
Nemchik:1998xb,Hufner:2000jb,Hoyer:2000fq}.
The cross section for \psits\ should be suppressed with respect to 
that for \jpsi, the dependence on \wgp\ should be somewhat steeper
than that of the \jpsi\ photoproduction cross section and
the $t$-dependence of \psits\ production 
should be similar or somewhat shallower than that of \jpsi\ production.
These predictions take into account the \psip\ meson wavefunction which
is different from that of the \jpsi\ meson in two respects.
It has a larger expectation value for the radius than the ground state 
and it has a node (see e.g.\,\cite{Hufner:2000jb}). 
The cross section for elastic \psits\ production has been measured previously
at HERA energies and the 
resulting
ratio $\sigma_{\psi(2S)}/\sigma_{J/\psi} = 0.150\pm0.027(stat)\pm0.018(syst)\pm0.011(BR)$
\cite{Adloff:1998yv} verifies the prediction of a
suppression with respect to 
\jpsi\ production. This and the other predictions mentioned above
are addressed in this paper. 

In a geometric interpretation
one expects the $t$-dependences, parametrised as 
$\propto e^{bt}$, to reflect the sizes of the interacting objects,
similarly to the hadroproduction case. Vector mesons with large
radii should have larger values of $b$ 
than vector mesons with small radii. 
This has indeed been confirmed in
photoproduction in the HERA energy range, where e.g.\,$b_{\rho^0}>b_{J/\psi}$ 
was measured \cite{Derrick:1995vq,Aid:1996bs}.  
Since the radius of the \psits\ meson is approximately a factor of
two larger than that of the 
\jpsi\ meson
one might naively expect a steeper $t$ dependence. 
In QCD a different result is obtained.
Due to cancellations in the contributions to
the production amplitude from $c\overline{c}$-quark dipoles with 
sizes above and below the node of the \psits\ wavefunction, 
the $t$-dependence of elastic \psip\ production
has been
predicted to be 
slightly shallower than 
that of the $J/\psi$ meson \cite{Nemchik:1998xb}. 

Calculations of the $t$-dependences for 
light and heavy vector meson 
production 
also exist \cite{Ryskin:1999qz} using an additive quark model
ansatz.
These calculations are able to reproduce the measurements of the
$t$-dependences at small $|t|$ for many elastic and proton dissociative
processes.

\section{Data Analysis}
\label{sec:analysis}

\subsection{Kinematics}
\label{sec:kinematics}

\noindent
The kinematics of the process $ep\rightarrow e\psi Y$ are described by 
the following variables: the square of the $ep$ center-of-mass energy $s=(p+k)^2$;
the negative four-momentum transfer squared at the lepton
vertex $Q^2=-q^2=-(k-k')^2$;
the four-momentum transfer squared at the proton
vertex $t=(p-p')^2$ and the scaled energy transfer $y=(p \cdot q)/(p \cdot k)$.
The four-momenta $k$, $k'$, $p$, $p'$ and $q$ refer to the
incident and scattered lepton, the incoming and outgoing proton or 
excited state $Y$
and the exchanged photon, respectively.
The elasticity $z$ of the meson production process
is defined as $z=(p \cdot p_{\psi})/(p \cdot q)$
where $p_{\psi}$ denotes the four-momentum of the produced vector 
meson. In the proton rest frame, $z$ describes
the fractional photon energy transferred to the
vector meson. 
It is related to $M_Y$ by $z \simeq 1 - (M_Y^2 - m_p^2 - t)/W_{\gamma p}^2$.
For elastic events at low $|t|$, in which the proton stays intact, $z \simeq 1$. 

\noindent 
In the photoproduction domain, i.e.\,at $Q^2 \rightarrow 0 $, 
the incoming lepton 
is scattered at small angles and below $Q^2 \sim 1$ GeV$^2$ it is
not observed in the central detector.
In the limit of photoproduction $W^2_{\gamma p} = (p+q)^2 $ is given by $ys$
where $y=(E-p_z)_{\psi}/(2E_e)$. Here $E$ and $p_z$ denote the total energy and 
momentum component of the vector meson 
parallel to the proton beam direction\footnote{The 
coordinate 
system of H1 defines the positive $z$-axis to be in the direction of the
proton beam. The polar angle $\theta$ is then defined with
respect to this axis.} of the vector meson and
$E_e$ is the energy of the incident beam lepton.
For the measurement of the elasticity the relation 
$z \simeq (E-p_z)_{\psi}/\Sigma(E-p_z)$ is used where $\Sigma(E-p_z)$
includes all measured final state particles.
The variable $t$ is approximated by the negative transverse momentum squared of
the vector meson, i.e.\,$t\simeq -p_{t,\psi}^2$. 

\subsection{Detector and Data Selection}
\label{sec:detselect}

\noindent
The H1 Detector is described in detail elsewhere \cite{Abt:1997xv}.
For this analysis the central and forward tracking detectors,
consisting of a system of drift and proportional chambers with
a polar angular coverage between $7^{\circ}$ and $165^{\circ}$,
are used for the detection of charged decay particles.
In addition the main and backward calorimeters, covering the
polar angular regions $4^{\circ}-153^{\circ}$ and 
$153^{\circ}-177.5^{\circ}$, respectively,
are used for lepton identification and for the determination of the
event kinematics \cite{Abt:1997xv,Nicholls:1996di}. 
The instrumented iron return yoke of the solenoidal magnet
($4^{\circ}<\theta < 171^{\circ}$)
which surrounds the central H1 detector supplements the 
calorimeters in the identification of muons.
Triggers based on lepton and track signatures are used to collect the events.

\noindent
The analysis presented here is based on data corresponding to an
integrated luminosity of $ 77 \:\mbox{pb}^{-1}$.
The data were taken in the years 1996 to 2000. Until 1997 
HERA was operated with positrons 
of energy $27.5\,\gev$ and protons of $820\,\gev$, while afterwards
the proton energy was $920\,\gev$ and both electrons and positrons were used.  
Events from $\psi(2S)$ and $J/\psi$ production
are selected using the direct decays 
into two leptons $\mu^+\mu^-$ or $e^+e^-$ and the $\psi(2S)$ is also reconstructed 
via the cascade into a $J/\psi$ and 
two charged pions with subsequent decay of the $J/\psi$ into two leptons.
The data selection resembles closely the procedure 
used in \cite{Adloff:1998yv,Adloff:2000vm}. 
Events with
a scattered lepton candidate detected in the calorimeters
with an energy deposit of more than $8\,\GeV$ are rejected. The accepted 
photoproduction event sample covers the range $Q^2 \lesssim 1\, \GeV^2$ with an
average $\langle Q^2\rangle \sim 0.055$ GeV$^2$ as determined from 
the Monte Carlo simulation.
\begin{figure}[t]
\center
\setlength{\unitlength}{1cm}
\begin{picture}(15.0,9.2)
\epsfig{file=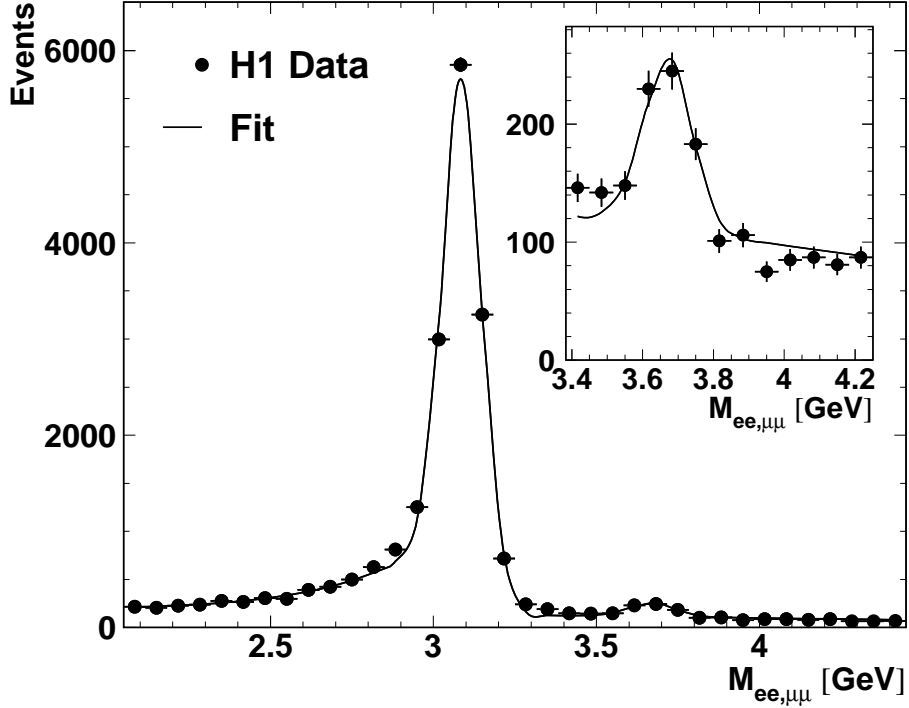}
\end{picture}
\caption{Mass spectrum ($|t|>0.07$ GeV$^2$) for the direct decay
  channel into leptons ($\mu^+\mu^-$ and $e^+e^-$) after the final selection. 
  The insert shows the mass distribution
  restricted to the $\psi(2S)$ region.
  The curve shows the result of a fit combining
  Gaussian distributions for the $J/\psi$ and \psip\ signals,
  an exponential parametrisation of the radiative tail in the electron 
  decay channel for the $J/\psi$ and a linear background.}
\label{fig:signal} 
\end{figure}

Vector meson decays into lepton pairs are selected by requiring exactly two 
tracks in the central tracking chambers, 
each with a transverse momentum greater than $0.8 \GeV$ 
in a polar angular 
range\footnote{For the data sample recorded in the year 2000, 
the polar angle of the decay electrons was required to 
be in the range $30^\circ < \theta < 150^\circ$.} 
of $20^\circ < \theta < 160^\circ$.
The tracks must define an event vertex in the $ep$ interaction region. 
Both tracks are required to
satisfy 
lepton identification requirements.
The decay electrons are identified using the electromagnetic section of 
the calorimeters and energy loss $dE/dx$ in the tracking chambers.
Muons are identified in the instrumented iron return yoke 
or as minimum ionizing particles in the main calorimeter. 
Cosmic ray muon events are rejected using a track 
acollinearity requirement.
For the selection of $\psi(2S)$ events via the cascade decay, 
$\psi(2S) \rightarrow ( J/\psi \rightarrow \ell^+\ell^-) \pi^+ \pi^-$,
events with exactly four tracks are selected. In addition to two
lepton candidate tracks, which are identified as described above, exactly 
two further central tracks with opposite charge are required, each with 
a transverse momentum 
of at least $0.12$ GeV.
The invariant mass of the two leptons is restricted to 
2.4 (2.9) $< M_{\ell\ell} <$ 3.3 GeV for cascade decays with 
$J/\psi \rightarrow e^+e^-$ ($J/\psi \rightarrow \mu^+\mu^-$), respectively.

\begin{figure}[t]
\center
\setlength{\unitlength}{1cm}
\begin{picture}(15.0,4.)
\epsfig{file=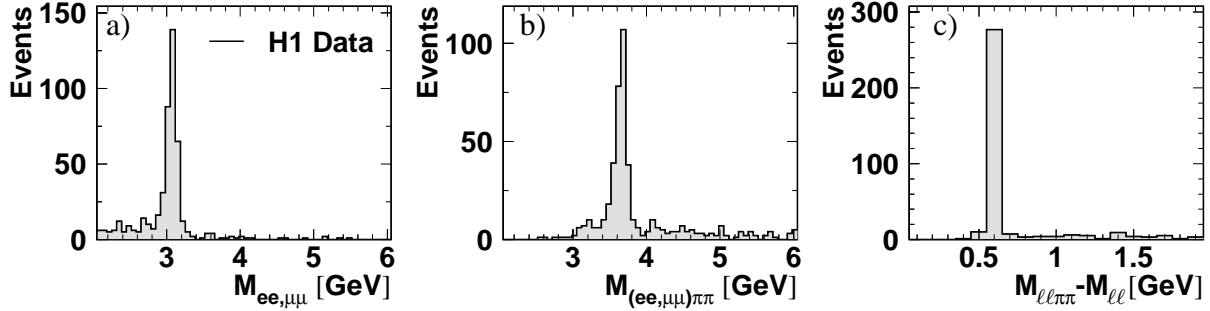}
\end{picture}
\caption{Mass spectra for the selected $\ell\ell\pi\pi$ sample; 
  a) and b) show the $\ell\ell$ and $\ell\ell\pi\pi$ invariant mass 
  distributions; c) shows the mass difference 
  $M_{\ell\ell\pi\pi}-M_{\ell\ell}$ 
  when selecting the $\ell\ell$ pairs reconstructed in the $J/\psi$ mass window.} 
\label{fig:psipsignal} 
\end{figure}

For 
the selection of the final diffractive (i.e. the sum of elastic and proton 
dissociative)
event samples, further requirements are
applied which ensure that the detector is essentially ``empty'' apart from
the \jpsi\ and \psip\ decay products.
These requirements 
and the method used to separate the elastic and
proton dissociative contributions
are discussed in detail in section~\ref{sec:separation}.
Figure~\ref{fig:signal} shows the di-lepton mass distribution 
for the diffractive sample in the mass region of the 
\jpsi\ and \psits\ mesons. 
The insert shows the same mass distribution restricted to the 
$\psi(2S)$ region revealing a clear signal
of $307 \pm  27$ events 
above the non-resonant background.
In decays with electrons the signals have a tail to lower masses due to the presence 
of radiative decays and radiation in the 
material of the detector. 
The non-resonant background is dominated
by the process $\gamma\gamma \rightarrow \ell^+\ell^-$ in which the
two photons couple to the beam lepton and to the proton, respectively. 
For the direct decays of \psits\ mesons 
the non-resonant background fraction is large. It is smaller for the 
\jpsi$\ra \ell^+\ell^-$ events and for the cascade decays of \psits\ mesons.
Figure~\ref{fig:psipsignal}a and b
show the $\ell\ell$ and $\ell\ell\pi\pi$ mass distributions for the 4-prong sample. 
In Fig.~\ref{fig:psipsignal}c the mass difference $M_{\ell\ell\pi\pi}-M_{\ell\ell}$ is shown.
A signal of 278 events in the range 
$|M_{\ell\ell\pi\pi}-M_{\ell\ell}-0.59~\rm GeV|< 0.06$ GeV
is seen with negligible background.

\subsection{Monte Carlo Models and Acceptances}

\noindent 
The acceptances and efficiencies for triggering, track reconstruction,
event selection and lepton identification are calculated using
Monte Carlo simulations. The \jpsi\ and \psits\ samples are generated using
the program DIFFVM \cite{diffvm} and are passed 
through a detailed simulation of the
detector response based on the GEANT program \cite{Brun:1987ma} and the same
reconstruction software as was used for the data.
DIFFVM generates events according to the cross section dependences $d\sigma/dt
\propto W^{4\epsilon}_{\gamma p}e^{bt}$ for elastic photoproduction of
charmonium. For production with proton dissociation 
$d^2\sigma/dt dM_Y^2 \propto W^{4\epsilon}_{\gamma
p}e^{b't}M_Y^{\beta}$ is used.
The parameters were chosen such that
the main features of $J/\psi$ and $\psi(2S)$ production are described:
$b=4.8$ GeV$^{-2}$; $b'=1.6$ GeV$^{-2}$; $\beta=-2.16$ and $4\epsilon=0.96$. 
Possible deviations 
from these parametrisations are taken into account in the systematic error
analysis.
The decay angular distributions of charmonium decaying directly into
two leptons are simulated assuming $s$-channel helicity conservation.
For background estimation the generators LPAIR \cite{Baranov:1991yq}
and GRAPE \cite{Abe:2001cv} 
are used, which simulate the process $\gamma \gamma \rightarrow \ell^+\ell^-$.
The contributions of radiative decays to \jpsi\ or \psits$\ra e^+e^-$ 
are estimated with the generator PHOTOS \cite{Barberio:1994qi}.

\begin{figure}[t]
\center
\setlength{\unitlength}{1cm}
\begin{picture}(15.0,8.5)
\put(1.5,-0.4){\epsfig{file=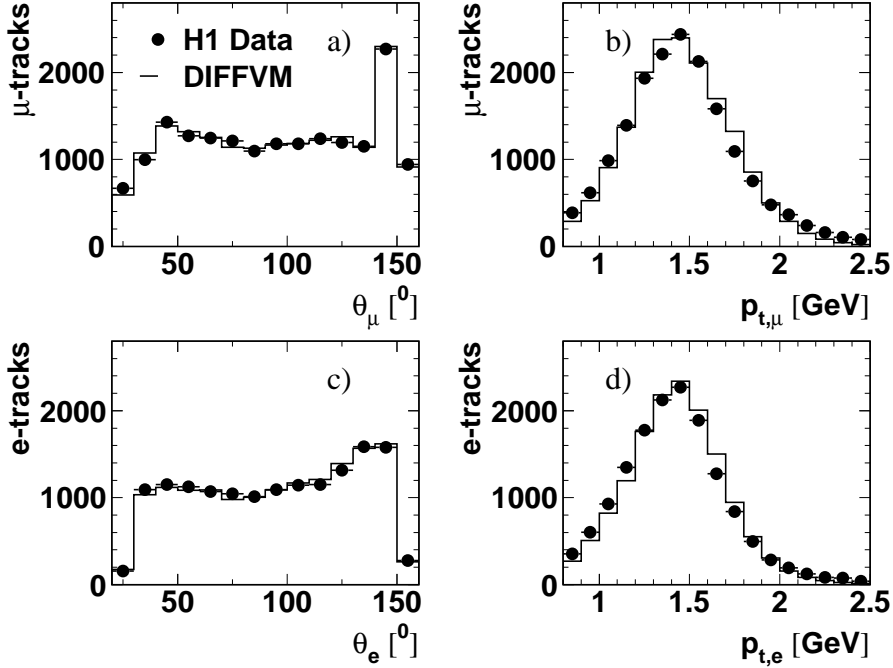}}
\end{picture}
\caption{Distributions for the decays of \jpsi\ mesons 
into muons (a,b) and electrons (c,d). The polar angles (a,c) and 
the transverse momenta (b,d) of the decay 
leptons are shown for the data (points) 
and the DIFFVM Monte Carlo simulation (histograms). The Monte Carlo 
simulation is 
normalised to the number of events in the data.}
\label{fig:checkdirect} 
\end{figure}

\begin{figure}[t]
\center
\setlength{\unitlength}{1cm}
\begin{picture}(15.0,13.5)
\put(1.5,-0.6){\epsfig{file=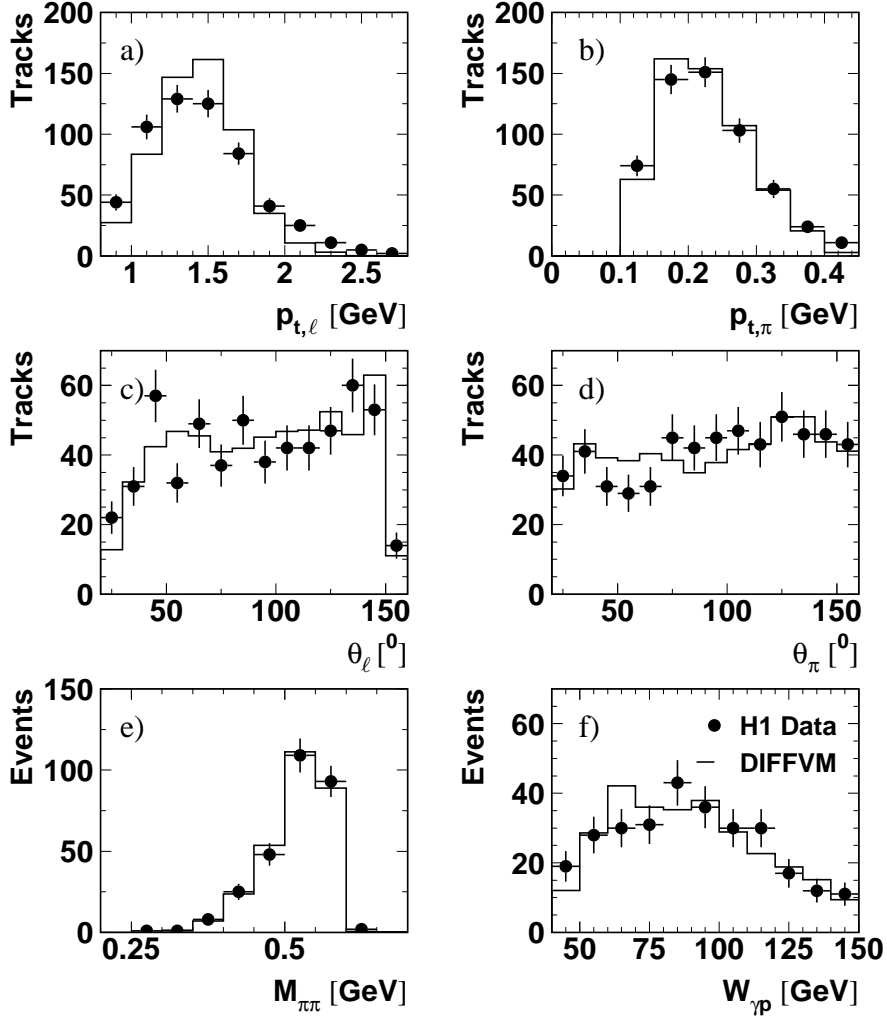}}
\end{picture}
\caption{Distributions for the cascade decays of $\psi(2S)$ mesons: 
a) the transverse momentum of the
decay leptons and b) the decay pions;
c) 
the polar angle of the decay leptons and d) the decay pions;
e) the two-pion invariant mass and f) 
the photon proton center-of-mass energy \wgp. \
The points show the data. The histograms show the DIFFVM Monte Carlo
simulation normalised to the number of events in the data 
after event reweighting (see text).}
\label{fig:checkcascade} 
\end{figure}

Detailed comparisons between the
simulation and the data are facilitated by the large 
$J/\psi$ data sample in which the background fraction is small.
Figure~\ref{fig:checkdirect} shows
the transverse momentum and polar angular distributions
of the tracks in the direct decays of $J/\psi$ into muons 
and electrons both for data and simulation. There is good agreement.
Figure~\ref{fig:checkcascade} shows that the simulation
for the cascade decays of $\psi(2S)$ adequately describes the data.
The simulation generates a flat 
$\pi\pi$-mass distribution in the available phase space,
ignoring the dependence arising from the matrix element.
To correct for this the simulation was
reweighted using an event weight $\propto (M_{\pi\pi}^2 - 4m^2_\pi)^2$ 
as determined in \cite{Pham:1976zq}. This results in a good description
of the $M_{\pi\pi}$ spectrum (Fig.~\ref{fig:checkcascade}e).
 
The detector simulation has been checked
extensively and separately for each data taking period and decay lepton type 
using independent data samples and was adjusted where necessary.
Typical trigger efficiencies are 45\% for the muon channel and
55\% for the electron channel. The efficiency for 
identifying a muon (electron) is typically 75\% (85\%).
Together with
the geometric detector acceptance the overall efficiency
varies between 5 and 10\% with \wgp. In the range studied here
the variation of the overall efficiency with $|t|$ is small.
The resolution in $t$ for the muon decay channel is typically 0.035 GeV$^2$ 
and increases to $0.060$ \GeV$^2$ at $|t|=1.2 \GeV^2$.
For the electron decay channel the resolution is roughly 
$15\%$ worse than this due to bremsstrahlung.

\subsection{Separation of Elastic and Proton Dissociative Processes}
\label{sec:separation}

For the analysis of the $t$-dependence it is necessary 
to separate elastic events from those with dissociation
of the proton into a small mass system.
To tag events with proton dissociation 
the forward section of the
calorimeter ($\theta< 10^\circ$), the
Proton Remnant Tagger ($0.06^{\degree}<\theta<0.26^{\degree}$)…
and the Forward Muon Detector 
($3^{\degree}<\theta<17^{\degree}$) are used.
With these forward detectors, dissociated proton states
with masses $M_Y \gtrsim 1.6\, \GeV$ can be tagged
\cite{Ahmed:1995ns}. 
In the following, events with (without) signals in the forward 
detectors are called `tagged' (`untagged'), respectively.

In the `untagged' sample no additional tracks other than those from 
the $J/\psi$ or \psip\ decay products are allowed.
In the `tagged' sample at most one additional track measured in the forward
tracking chambers in the polar angular region below 
$10^{\circ}$ is allowed and an elasticity $z > 0.95$ is required
in order to ensure that the event sample is diffractive.

\noindent
The elastic and proton dissociative contributions to the diffractive sample
are extracted from the untagged and tagged event samples respectively,
taking the background admixtures
from proton dissociative events in the untagged sample 
($\sim 15\% 
$)
and elastic events in the tagged sample ($\sim 10\%
$) into account.
The admixtures are due to the limited angular coverage of the forward detectors,
their inefficiencies and noise fluctuations. 

\begin{table}[ht]
\begin{center}
    \begin{tabular}{|l|cc|cc|cc|}
      \hline
       & \multicolumn{2}{|c|}{ ~~~$\psi(2S) \rightarrow \ell\ell $~~~ } &
         \multicolumn{2}{|c|}{ $ \psi(2S) \rightarrow \ell\ell\pi\pi $ } &
         \multicolumn{2}{|c|}{ $~~~J/\psi \rightarrow \ell\ell ~~~$ } \\
                   & untagged & tagged &
                     untagged & tagged &
                     untagged & tagged \\
      \hline
      Elastic                &  39\% &  \phantom{0}6\%& 85\%& 10\% & 80\% &  \phantom{0}9\% \\
      Proton Dissociative    &  \phantom{0}6\% & 54\%& 15\%& 90\% & 14\% & 85\% \\
      Non-resonant Bg.       &  55\% & 40\%& -- & -- &  \phantom{0}6\% &  \phantom{0}6\% \\
      \hline
    \end{tabular}
    \end{center}
    \caption{Estimated contributions to the tagged and untagged
    samples for the three different charmonium decays and the non-resonant
    backgrounds (see text).}
    \label{tab:relfracs}
\end{table}

In Table~\ref{tab:relfracs} the estimated relative signal and background contributions
are listed for the tagged and untagged samples of the different decay channels.
The non-resonant background in the mass window of $\pm 150$ MeV around the
nominal charmonium masses is determined from the side-band events. It is found
to be negligible for the \psip\ cascade decay samples. 
The elastic and proton dissociative contributions in Table~\ref{tab:relfracs}
are determined using a Monte Carlo simulation of the forward region
of the detector and the
beamline. In the simulation the relative contributions
from the proton dissociative and elastic channels
are adjusted to reproduce the distributions measured in the forward detectors.
The uncertainty on the determination of the admixtures is evaluated by variation
of the simulated forward detector efficiencies and of the simulated cross section 
dependence on the mass $M_Y$ of the dissociated proton system \cite{schmidt}. 
The resulting
uncertainty on the final cross sections is 6\%.
Within experimental errors, equal cross sections for elastic
and proton dissociative
photoproduction of charmonium are found for the measured $M_Y$ and $t$ regions,
which is
in agreement with previous results \cite{Aid:1996dn,Adloff:1998yv,Adloff:2000vm}.

The fraction of erroneously tagged events in the elastic sample increases
with $|t|$ and, with the present analysis method, 
no distinction between elastic and proton dissociative events can be 
made for $|t| \gtrsim 1$ GeV$^2$.
However, since the $t$-dependence of the elastic channel is much
steeper than that for proton dissociation, the elastic contribution
at $|t| \gtrsim 1$ GeV$^2$ can safely be neglected.

\subsection{\boldmath \psits\ and \jpsi\ Signal Determination}
\label{sec:signal}

For the measurement of the \wgp\ dependence
the numbers of signal events are determined for the diffractive samples 
in four bins of \wgp.
For events with direct decays into muons a simultaneous fit is used
in which the \psip\ or \jpsi\ signals are parametrised by Gaussian 
distributions and the non-resonant background follows a linear dependence.
In an alternative method the non-resonant background is estimated using 
the generator LPAIR \cite{Baranov:1991yq} and is subtracted from
the number of signal events, counted in mass windows
of $\pm 150$ MeV width around the nominal $J/\psi$ and \psip\ masses.
For the number of $J/\psi$ signal events agreement 
within 2\% between the two methods is found.
For the direct $\psi(2S)$ decays into muons the number
of signal events obtained from the fits 
differs by up to 10\% from that obtained when using
LPAIR to describe the background. 
These differences are taken as systematic errors.

For the direct decays of $\psi(2S)$ and $J/\psi$ 
into electrons the number of signal events
is determined using a fit to the
mass spectrum with Gaussians for the $J/\psi$ and $\psi(2S)$ signals,
an exponential distribution for the radiative tails and a linear dependence
for the non-resonant background.
The systematic error in this method is estimated by
comparing the result for the non-resonant background
with that of the Monte Carlo generator GRAPE \cite{Abe:2001cv}. 
For $J/\psi$ decays into electrons the resulting systematic error is 5\%. 
For the direct $\psi(2S)$ decays into electrons
the uncertainty on the signal determination 
is estimated to be 15\%, which is 
larger than that for \jpsi\ due to the larger background fraction.

For the cascade decays the non-resonant background is very small and
all events found in the range
$|M_{\ell\ell\pi\pi}-M_{\ell\ell}-0.59~\rm GeV|< 0.06$ GeV are assigned to
the signal (see Fig.~\ref{fig:psipsignal}c).
The number of $J/\psi$ events is corrected for the fraction of 
$\psi(2S)$ events with decays into a $J/\psi$ and neutral 
particles. 
This correction is estimated to 
be $(3.4\pm0.9)\%
$ based on the results of this analysis (see below). 
The contribution to the $\psi(2S)$ sample
from decays of $\psi(3S)$ and higher excited states
is expected to be small and is neglected.

\section{Results}

The results are given 
for an average $\langle Q^2\rangle \sim 0.055$ GeV$^2$ 
and cover an energy range $40<\wgpw <150\, \GeV$. 
For the proton dissociative channel the kinematic region is 
restricted to $|t|<5$ GeV$^2$ and $(M_Y/W_{\gamma p})^2<0.05$.

\subsection{Energy Dependence of \boldmath $\sigma(\psi(2S))/\sigma(J/\psi)$ }
\label{sec:psip}

The ratio of \psip\ to \jpsi\ cross sections 
$R=\sigma(\psi(2S))/\sigma(J/\psi)$ is measured as a function of \wgp.
The total diffractive 
$\psi(2S)$ and $J/\psi$ samples, i.e.\,the sum of the tagged and the 
untagged samples as defined in section \ref{sec:separation}, are used in order
to minimise the systematic and statistical errors. 
The experimental signatures of the leptonic decay channels of 
$\psi(2S)$ and \jpsi\ mesons are very similar and the large 
sample of \jpsi\ mesons has been used to support the study of the 
experimental systematics for the \psip\ measurement.
In the context of this analysis a complete measurement of the $J/\psi$ 
cross section was performed \cite{schmidt},
yielding a \wgp\ dependence which is consistent 
with the results of our earlier publication \cite{Adloff:2000vm}.

The cross section ratio is obtained by taking the ratio of the corrected
numbers of $\psi(2S)$ events to the corrected number of $J/\psi$ events 
in each \psip\ decay channel separately.
Each of the samples is corrected for its specific trigger 
efficiency, event selection efficiency
and the \psip\ and \jpsi\ branching ratios\footnote{The branching 
ratios for the \psip\
are $(0.79\pm0.05)\%$, $(0.77\pm0.17)\%$ and $(34.8\pm2.8)\%$ for the decays 
to $e^+e^-$, $\mu^+\mu^-$ and $J/\psi\pi^+\pi^-$, respectively.
For the $J/\psi$ decays into muons and electrons 
the branching ratios are $(5.88\pm0.10)\%$ and $(5.93\pm0.10)\%$, 
respectively \cite{Groom:2000in}.}.
In the evaluation of the systematic error on the cross section ratio
the errors on the integrated luminosity, 
the trigger efficiency, 
the lepton identification and track reconstruction efficiencies 
and the detector acceptance
are found to be highly correlated between the $J/\psi$ and the \psip\ samples
and therefore largely cancel. 
The remaining 
errors from these sources amount to 5\% in total.
For the direct decays the uncertainty in the determination
of the number of \psip\ signal events 
is the dominant systematic error contributing 10\% for decays to $\mu^+\mu^-$ 
and 15\% for decays to $e^+e^-$ (see section~\ref{sec:signal}).
For the cascade decays the uncertainty in the pion track reconstruction 
efficiency of 12\% dominates the systematic error.

\begin{figure}[tp]
\center
\setlength{\unitlength}{1cm}
\begin{picture}(15.0,7.8)
\put(1.5,-0.6){\epsfig{file=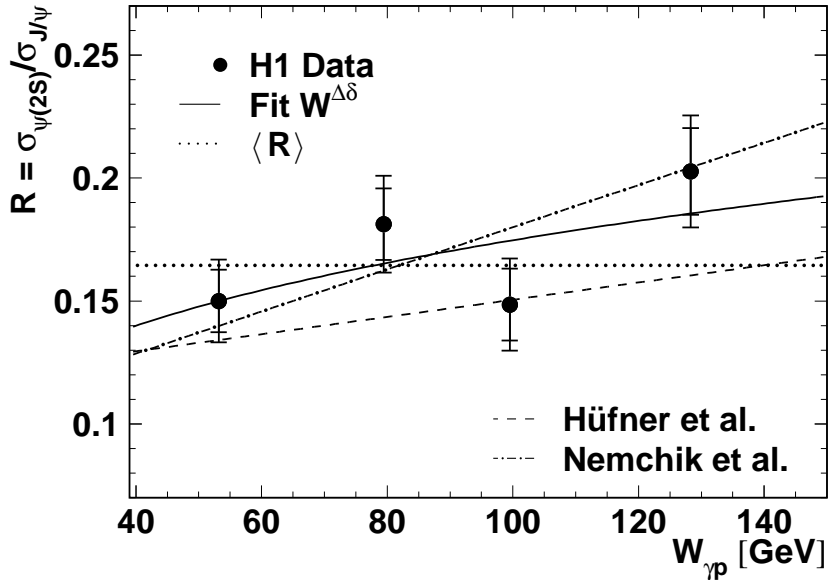}}
\end{picture}
\caption{The ratio 
$R(W_{\gamma p})=\sigma_{\psi(2S)}/\sigma_{J/\psi}$ 
for events with $z>0.95$. The inner error bars show the statistical error.
The outer error bars show the statistical and systematic
errors added in quadrature. An additional normalisation uncertainty 
of 0.007 due
to the errors on the branching ratios is not shown.
A fit $R \propto (W_{\gamma p}/90~{\rm GeV})^{\Delta\delta}$
with
a value for ${\Delta\delta}$ of 0.24 (solid line) 
and predictions from \cite{Nemchik:1998xb} (dashed-dotted line) 
and \cite{Hufner:2000jb} (dashed line) are also shown.}
\label{fig:psi2-ratiow} 
\end{figure}

\noindent
Figure~\ref{fig:psi2-ratiow} and Table~\ref{tab:ratiow}
show the measured cross section ratio as a function of $W_{\gamma p}$.
In the figure the inner error bars reflect the statistical errors,
the outer error bars show the statistical and the systematic
errors added in quadrature, excluding the uncertainty from 
the branching ratios. 
The overall ratio, shown in Fig.~\ref{fig:psi2-ratiow} as a
dotted line, yields the value
$$
R=0.166\pm0.007(stat.)\pm0.008(sys.)\pm0.007(BR),
$$
in good agreement with an earlier measurement \cite{Adloff:1998yv}.
This result is obtained by averaging over the ratios
for the different decay channels and integrating over the four bins
in $W_{\gamma p}$, taking proper account of the correlations in the branching
ratio errors.
The solid line shows a fit of the 
form $R(W_{\gamma p})\propto W_{\gamma p}^{\Delta\delta}$ to the 
data points yielding a value of
$\Delta\delta=0.24\pm0.17$ with $\chi^2=3.1$ for 2 degrees of freedom,
where the error includes systematic and statistical uncertainties 
added in quadrature. 
The energy dependence of the diffractive \psits\ photoproduction cross section 
is thus similar or possibly slightly steeper than that for \jpsi\ mesons.
The calculations of \cite{Nemchik:1998xb}
and of \cite{Hufner:2000jb}\footnote{The different parametrisations 
given in \cite{Hufner:2000jb} vary
considerably in normalisation. In Fig.~\ref{fig:psi2-ratiow}
the parametrisation referred to in \cite{Hufner:2000jb} as 'GBW(BT)' 
is shown.} which 
are valid for elastic photoproduction
are also shown. Both groups predict a somewhat 
steeper energy dependence for the $\psi(2S)$ than for the $J/\psi$ meson,
compatible with the data in both slope and normalisation. 

\subsection{{\boldmath $t$}-Dependences of the Elastic and 
Proton Dissociative Channels}
\label{sec:tdep}

\begin{figure}[p]
\center
\setlength{\unitlength}{1cm}
\begin{picture}(15.0,18.0)
\put(-0.6,-0.3){\epsfig{file=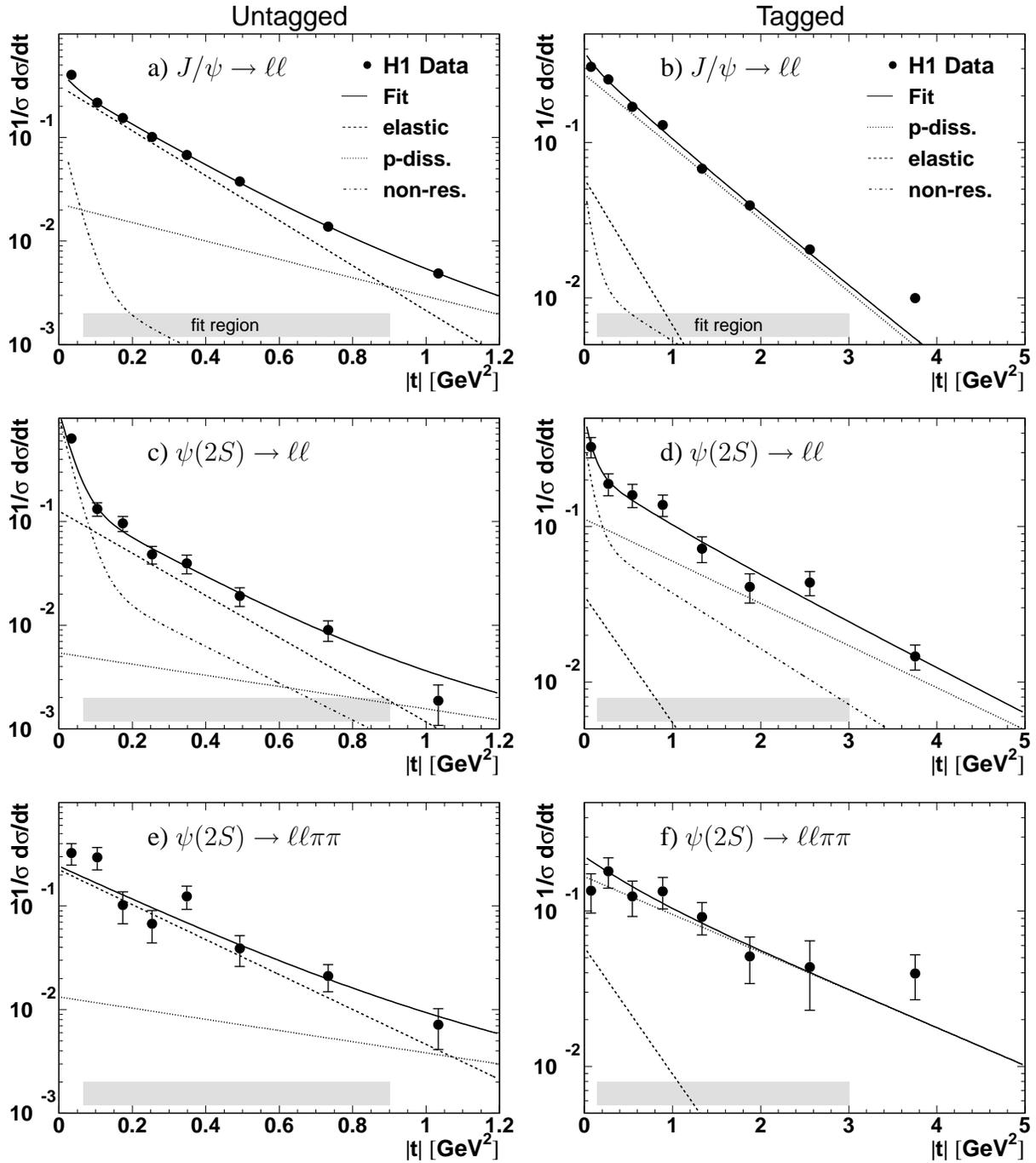}}
\end{picture}
\caption{Normalised differential cross sections $1/\sigma \cdot d\sigma/dt$
as a function of $|t|$ for events a) in the $J/\psi$ mass window without
a signal in the forward detectors (untagged), b) in the $J/\psi$ mass window with 
a signal in the forward detectors (tagged), c) untagged and d) 
tagged events in the $\psi(2S)$ mass window for direct decays to lepton pairs, 
e) untagged and f) tagged $\psi(2S)$ events with cascade decays.
The solid lines show the results of the fits described in the text. 
The dashed (dotted)
curves show the contributions from the elastic (proton
dissociative) processes, respectively.
For the direct decays into leptons (a-d) 
the contributions from the non-resonant background (dashed-dotted curves)
are also shown. The shaded bands indicate the fit regions.}
\label{fig:psi2-t} 
\end{figure}

In order to study the $t$-dependences, the data are divided into the tagged
and untagged samples as described in section~\ref{sec:separation}.
Figure~\ref{fig:psi2-t} shows the normalised differential cross sections
$1/\sigma \cdot d\sigma/dt$ for events in mass windows
of $\pm$ 150 MeV around the nominal \jpsi\ and \psip\ masses
for each of the three samples $J/\psi$, $\psi(2S)$ with direct and 
$\psi(2S)$ with cascade decays.
The data samples consist of events with decays into both muons and electrons
and are corrected for efficiencies but not for remaining backgrounds.
The untagged (tagged) cross sections 
are displayed in the left (right) column of the figure.
The main contribution to the untagged cross section is from 
the elastic diffractive process 
with contaminations from 
the untagged proton dissociative process  and from
non-resonant background.
The main contribution to the tagged cross section is from the proton dissociative 
process, with contaminations from the non-resonant
background and from the elastic
process.

The method used to analyse the $t$-dependence
is optimised for samples with large non-resonant 
background, i.e.\,the direct \psip\ decays.
The same method is used for the analysis
of the $t$-dependences of the $J/\psi$ samples and the
$\psi(2S)$ samples with cascade decays, for consistency.
The $t$-dependences of the elastic and the proton dissociative 
contributions are described
by single exponential distributions $e^{bt}$. 
The slope parameters $b$ are extracted using an iterative procedure of combined fits
to the tagged and untagged samples.
The non-resonant backgrounds are described by sums of two exponential
functions. The $t$-dependence and relative normalisation of the 
non-resonant background are determined 
from an analysis of the side-band events in the range $4 < M_{\ell\ell} < 8$ GeV,
and are fixed in the fit procedure. 
The observed behaviour is reproduced by the LPAIR and GRAPE generators.

  
\noindent 
In the iteration procedure
the relative fractions of the background contributions
are fixed to the numbers given
in section \ref{sec:separation}.
In the first step a fit is performed for the proton dissociative
slope parameter in the tagged sample, neglecting the elastic contamination. 
This estimate is then used to compute the 
proton dissociative contamination in the untagged sample.
The Monte Carlo simulation is used in this calculation 
to account for the $t$-dependence
of the probability of failing to tag proton dissociation events.
Secondly, a fit to the untagged sample 
is performed to extract the slope parameter of the 
elastic channel, correcting for the untagged proton dissociative contribution.
This measurement of the elastic channel is then used to estimate 
the elastic contribution in the tagged sample and to 
improve the measurement of the proton dissociative process obtained in the 
first step.
The fit procedure converges after one iteration since
the elastic contamination in the tagged sample
is small (see Fig.~\ref{fig:psi2-t}).   
The elastic slope parameters $b_{el}$ 
are extracted for each of the three untagged samples
in the 
region of $0.07 < |t| < 0.9$ GeV$^2$ excluding the regions
with large non-resonant background (lowest $|t|$ bin) or large contribution 
from proton dissociation (highest $|t|$ bin).
Similarly, the proton dissociative slope parameters $b_{pd}$
are extracted from the tagged samples in the 
region of $0.15 < |t| < 3$ GeV$^2$. 
The lowest $|t|$ bin is excluded since for proton dissociation, 
the minimum kinematically allowed value of $|t|$ can be large 
for large masses $M_Y$. The highest $|t|$ bin is excluded because 
at large values of $|t|\gtrsim 3$ GeV$^2$, the data deviate 
from an exponential behaviour.
In Fig.~\ref{fig:psi2-t} the solid lines indicate 
the parametrisations of the data as obtained from the fits,
including the contributions from the signals and the backgrounds.

\noindent 
The systematic uncertainty on 
the $b$ values is estimated by varying the selection 
cuts and trigger conditions, the $|t|$-dependences of the efficiencies 
and acceptances 
and the admixtures of the background contributions 
(both relative normalisations and shapes)
resulting in the uncertainties given in Table~\ref{tab:bslopeerr}.
The dominant error source for the 
elastic slope parameter is the proton dissociative background. 
For the proton dissociative slope parameter 
the $t$-dependences of the trigger 
and event selection efficiencies are the dominant error sources.

\noindent 
From the untagged samples the fits yield the elastic slope parameters
$b_{el}^{\psi(2S)}=(4.31\pm0.57\pm0.46)$ GeV$^{-2}$ for the $\psi(2S)$ and
$b_{el}^{J/\psi}=(4.99\pm0.13\pm0.39)$ GeV$^{-2}$ for the $J/\psi$,
where the first error 
is statistical and the second error describes the systematic uncertainties.
The result for $b_{el}^{J/\psi}$ confirms the previous measurement 
\cite{Adloff:2000vm}.
From the tagged samples the proton dissociative parameters
$b_{pd}^{\psi(2S)}=(0.59\pm0.13\pm0.12)$ GeV$^{-2}$ 
for the $\psi(2S)$ 
and $b_{pd}^{J/\psi}=(1.07\pm0.03\pm0.11)$ GeV$^{-2}$ for the $J/\psi$ are
obtained. 
These results are listed in Table~\ref{tab:slopes}, together with
the results for separate fits to the
samples with muons or with electrons alone
and to the samples with direct and cascade $\psi(2S)$ decays.
Good agreement is observed
between the results from the different samples.

\noindent 
The measured slope parameters $b_{pd}$ with proton
dissociation are considerably smaller than in the elastic case. This
is in agreement with expectations \cite{Ryskin:1999qz}. 
The $t$-dependence of elastic $\psi(2S)$ photoproduction 
is similar to that of $J/\psi$.
This is in agreement with expectations in the additive quark model 
\cite{Ryskin:1999qz} and also within the colour dipole 
model \cite{Nemchik:1998xb}.
For proton dissociation a
somewhat shallower $t$-dependence is measured for the \psip\ than 
for the \jpsi. The difference between the proton dissociative
slope parameters of \jpsi\ and \psits\ mesons 
amounts to 2.3 standard deviations and can be interpreted
as an effect of the \psits\ wavefunction node on the $t$-dependence. 
In proton dissociation this is more visible than in the elastic process
due to the lower overall values of the slope parameter $b_{pd}$.

\section{Summary}
The $W_{\gamma p}$ and $t$-dependences of diffractive $\psi(2S)$ 
photoproduction have been measured for the first time.
For the total diffractive cross section ratio $R=\sigma_{\psi(2S)}/\sigma_{J/\psi}$ 
a value of
$$
R=0.166\pm0.007(stat.)\pm0.008(sys.)\pm0.007(BR)
$$
is obtained in the region 
$ 40 < W_{\gamma p} < 150 $ GeV and $Q^2 < 1$ GeV$^2$ consistent with
and improving the errors of our previous measurement \cite{Adloff:1998yv}.
The cross section ratio has been measured in four bins of $W_{\gamma p}$
and the result indicates
that the energy dependence of \psip\ photoproduction 
is similar or possibly somewhat steeper than that of
$J/\psi$ production. 
The data are well described by pQCD calculations \cite{Nemchik:1998xb,Hufner:2000jb}.

\noindent 
The $t$-dependences of elastic and proton dissociative charmonium production 
have been measured. 
The $t$-dependence of elastic $\psi(2S)$ meson production
is compatible with an exponential behaviour $\propto e^{bt}$
in the measured range of $|t|<0.9~\rm GeV^2$,
with $b_{el}^{\psi(2S)}=(4.31\pm0.57\pm0.46)$ GeV$^{-2}$. 
This value is similar to the result for \jpsi\ mesons, 
$b_{el}^{J/\psi}=(4.99\pm0.13\pm0.39)$ GeV$^{-2}$ and thus
confirms the predictions of \cite{Nemchik:1998xb,Hufner:2000jb}. 
The proton dissociative slope parameter has been determined to be
$b_{pd}^{\psi(2S)}=(0.59\pm0.13\pm0.12)$ GeV$^{-2}$ for the \psip\ meson, 
somewhat smaller than that for the \jpsi\ meson, which is measured to be
$b_{pd}^{J/\psi}=(1.07\pm0.03\pm0.11)$ GeV$^{-2}$.
These results are well described by pQCD calculations which take into
account the differences in the wavefunctions of the \jpsi\ and its radial
excitation.

\section*{Acknowledgments}

We are grateful to the HERA machine group whose outstanding
efforts have made and continue to make this experiment possible. 
We thank
the engineers and technicians for their work in constructing and now
maintaining the H1 detector, our funding agencies for 
financial support, the
DESY technical staff for continual assistance 
and the DESY directorate for the
hospitality which they extend to the non DESY 
members of the collaboration.

\newpage

\begin{table}[h]
\begin{center}
    \begin{tabular}{|c|c|c|}
      \hline
      \wgp\ Interval $[\rm GeV]$& $\langle W_{\gamma p} \rangle $ $[\rm GeV]$& $R(W_{\gamma p})$ \\
      \hline
       $ 40 -  70$      &    $  53.2   $        & $0.150 \pm 0.013 \pm 0.011$ \\
       $ 70 -  90$      &    $  79.4   $        & $0.181 \pm 0.015 \pm 0.013$ \\
       $ 90 - 110$      &    $  99.5   $        & $0.149 \pm 0.015 \pm 0.012$ \\
       $110 - 150$      &    $ 128.3   $        & $0.203 \pm 0.018 \pm 0.014$ \\
      \hline
    \end{tabular}
    \end{center}
    \caption{Results for the \psip\ to \jpsi\ photoproduction cross section 
    ratio $R=\sigma_{\psi(2S)}/\sigma_{J/\psi}$ in four bins of \wgp\
    together with the statistical and systematic errors.
    An additional normalisation uncertainty of 0.007 due to the 
    branching ratios is not included in the errors.}
    \label{tab:ratiow}
\end{table}

\begin{table}[h]
\begin{center}
    \begin{tabular}{|l|cc|cc|cc|}
      \hline
       & \multicolumn{2}{|c|}{ ~~~$\psi(2S) \rightarrow \ell\ell $~~~ } &
       \multicolumn{2}{|c|}{ $ \psi(2S) \rightarrow \ell\ell\pi\pi $ } &
        \multicolumn{2}{|c|}{ $~~~J/\psi \rightarrow \ell\ell ~~~$ } \\
               Systematic errors $[\rm GeV^{-2}]$    & ~$\Delta b_{el}$ & ~$\Delta b_{pd}$ &
                     ~$\Delta b_{el}$~ & ~$\Delta b_{pd}$~ &
                     ~$\Delta b_{el}$~ & ~$\Delta b_{pd}$~ \\
      \hline
      Trigger and selection efficiencies              &
                         $ 0.2\phantom{0} $ & $ 0.1\phantom{0} $ & $ 0.2\phantom{0} $ & $ 0.1\phantom{0} $ & $ 0.2\phantom{0} $ & $ 0.1\phantom{0} $ \\
      $t$-dependence of forward tagging    &
                         $ 0.2\phantom{0} $ & $ 0.03 $ & $ 0.2\phantom{0} $ & $ 0.03 $ & $ 0.2\phantom{0} $ & $ 0.03 $  \\
      Normalisation of non-res.~background &
                             $ 0.11 $ & $ 0.01 $
                           &     --      &    --
                           &   --     &  --      \\
      $t$-dependence of non-res.~background &
                             $ 0.05 $ & $ 0.04 $
                           &    --      &    --
                           &   --     &   --      \\
      Normalisation of $p$-diss.(elas.)~background &
                             $ 0.35 $ &   $ 0.01 $
                           & $ 0.35 $ &   $ 0.01 $
                           & $ 0.27 $ &   $ 0.01 $     \\
      $t$-dependence of $p$-diss.(elas.)~background &
                             $ 0.06 $ &   $ 0.05 $
                           & $ 0.06 $ &   $ 0.05 $
                           & $ 0.03 $ &   $ 0.03 $      \\
      \hline
      Total error          &
                             $ 0.47 $ & $ 0.12 $
                           & $ 0.45 $ & $ 0.12 $
                           & $ 0.39 $ & $ 0.11 $ \\
      \hline
    \end{tabular}
    \end{center}
    \caption{Table of systematic errors (in $\rm GeV^{-2}$) in the determination
    of the elastic and proton dissociative 
slope parameters $b$ of $\psi(2S)$ production with direct and cascade decays
    and $J/\psi$ production. }
    \label{tab:bslopeerr}
\end{table}

\begin{table}[h]
\begin{center}
\begin{tabular}{|l|c|c|}
           \hline
           & $b_{el}[\rm GeV^{-2}]$ & $b_{pd}[\rm GeV^{-2}]$ \\
           \hline
 $\psi(2S) \rightarrow \mu\mu$           &  $ 4.76\pm0.78~~~~~~~~~~~~ $ & $ 0.69\pm0.22~~~~~~~~~~~~  $    \\
 $\psi(2S) \rightarrow ee$               &  $ 3.51\pm2.44~~~~~~~~~~~~ $ & $ 0.42\pm0.35~~~~~~~~~~~~  $    \\
 $\psi(2S) \rightarrow \mu\mu\pi^+\pi^-$ &  $ 3.19\pm0.96~~~~~~~~~~~~ $ & $ 0.53\pm0.28~~~~~~~~~~~~  $  \\
 $\psi(2S) \rightarrow ee\pi^+\pi^-$     &  $ 5.91\pm2.74~~~~~~~~~~~~ $ & $ 0.57\pm0.26~~~~~~~~~~~~  $  \\
 $J/\psi \rightarrow \mu\mu$             &  $ 4.98\pm0.15~~~~~~~~~~~~ $ & $ 1.10\pm0.04~~~~~~~~~~~~  $      \\
 $J/\psi \rightarrow ee$                 &  $ 5.05\pm0.25~~~~~~~~~~~~ $ & $ 1.01\pm0.04~~~~~~~~~~~~  $      \\
           \hline
 $\psi(2S) \rightarrow \ell\ell$               &  $ 4.69\pm0.73\pm0.45 $ & $ 0.62\pm0.18\pm0.12  $         \\
 $\psi(2S) \rightarrow \ell\ell\pi^+\pi^-$     &  $ 3.88\pm0.92\pm0.47 $ & $ 0.56\pm0.19\pm0.12  $  \\
           \hline
 $\psi(2S) $                             &  $ 4.31\pm0.57\pm0.46 $ & $ 0.59\pm0.13\pm0.12  $ \\
 $J/\psi   $                             &  $ 4.99\pm0.13\pm0.39 $ & $ 1.07\pm0.03\pm0.11  $      \\
           \hline
\end{tabular}
\end{center}
\caption{Slope parameters for charmonium production
 for the elastic (left column) and proton dissociative
(right column) component. For the separate electron and muon channels 
only the statistical errors are given. For the results combining both
channels, both statistical and systematic errors are quoted.}
\label{tab:slopes}
\end{table}

\clearpage

\end{document}